\documentclass[conference]{IEEEtran}
\IEEEoverridecommandlockouts
\usepackage{cite}
\usepackage{amsmath,amssymb,amsfonts}
\usepackage{algorithmic}
\usepackage{graphicx}
\usepackage{textcomp}
\usepackage{xcolor}
\def\BibTeX{{\rm B\kern-.05em{\sc i\kern-.025em b}\kern-.08em
    T\kern-.1667em\lower.7ex\hbox{E}\kern-.125emX}}
\begin{document}

\title{Deep meta-learning for the selection of accurate ultrasound based breast mass classifier
}

\author{\IEEEauthorblockN{Michal Byra\IEEEauthorrefmark{1}\IEEEauthorrefmark{2}, Piotr Karwat\IEEEauthorrefmark{2}, Ivan Ryzhankow\IEEEauthorrefmark{2}, Piotr Komorowski\IEEEauthorrefmark{2}, \\ Ziemowit Klimonda\IEEEauthorrefmark{2}, Lukasz Fura\IEEEauthorrefmark{2}, Anna Pawlowska\IEEEauthorrefmark{2}, Norbert Zolek\IEEEauthorrefmark{2}, Jerzy Litniewski\IEEEauthorrefmark{2} }

\\

\IEEEauthorblockA{\IEEEauthorrefmark{2}Institute of Fundamental Technological Research, \\Polish Academy of Sciences, Warsaw, Poland}

\IEEEauthorblockA{\IEEEauthorrefmark{1}Corresponding author, e-mail: mbyra@ippt.pan.pl}
}

\maketitle

\begin{abstract}

Standard classification methods based on handcrafted morphological and
texture features have achieved good performance in breast mass differentiation in ultrasound (US). In comparison to deep neural networks, commonly perceived as
‘black-box’ models, classical techniques are based on features that have well-understood medical and physical interpretation. However, classifiers based on morphological features commonly underperform in the presence of the shadowing artifact and ill-defined mass borders, while texture based classifiers may fail when the US image is too noisy. Therefore, in practice it would be beneficial to select the classification method based on the appearance of the particular US image. In this work, we develop a deep meta-network that can automatically process input breast mass US images and recommend whether to apply the shape or texture based classifier for the breast mass differentiation. Our preliminary results demonstrate that meta-learning techniques can be used to improve the  performance of the standard classifiers based on handcrafted features. With the proposed meta-learning based approach, we achieved the area under the receiver operating characteristic curve of 0.95 and accuracy of 0.91. 

\end{abstract}


\section{Introduction}

Ultrasound (US) imaging is widely used to assess and diagnose breast cancer. Various classic and deep learning based methods have been proposed for breast mass classification in US \cite{houssein2021deep}. Deep neural networks have achieved excellent performance in breast mass differentiation, but are commonly perceived as black-box models difficult to incorporate into the clinical practice due to the lack of interpretability \cite{abdullah2021review}. In contrary, standard classification techniques based on handcrafted features have established medical and physical interpretation. Flores et al. conducted a detailed comparison between the texture and morphological features in the case of the breast mass differentiation in US \cite{flores2015improving}. Results based purely on classification metrics indicated that the morphological features are the better performing features for the mass differentiation. However, authors did not investigate why the shape or texture based methods failed to correctly classify specific US images. Standard classification methods may fail to produce accurate predictions for breast mass US images for several reasons.  For example, due to the shadowing artifact resulting in ill-defined mass borders the accurate estimation of the morphological features may be infeasible, see Fig. 1. Similarly, the texture based classifier may underperform when the US image is noisy or if the mass texture was adversely impacted by the US image processing algorithms~\cite{byra2019quantitative}.  Therefore, it should be expected in practice that the texture based methods will perform better on some cases while the shape based techniques will perform better on others. 

In this work, we propose a deep meta-learning based approach to the selection of the appropriate standard classification method for particular breast mass US image. In machine learning literature, meta-learning techniques have been used to recommend classification algorithms for specific tasks and datasets~\cite{khan2020literature}. Here, we develop a neural network that can automatically process input breast mass US image and recommend whether to apply the shape or texture based classifier for the analysis. By using meta-learning we aim to address the issues associated with the robustness of the standard classifiers. In our study, deep learning techniques are not used to directly classify breast masses, but to improve the performance of the standard methods based on well-understood handcrafted features. 

\begin{figure}[]
	\begin{center}
		\includegraphics[width=0.9\linewidth]{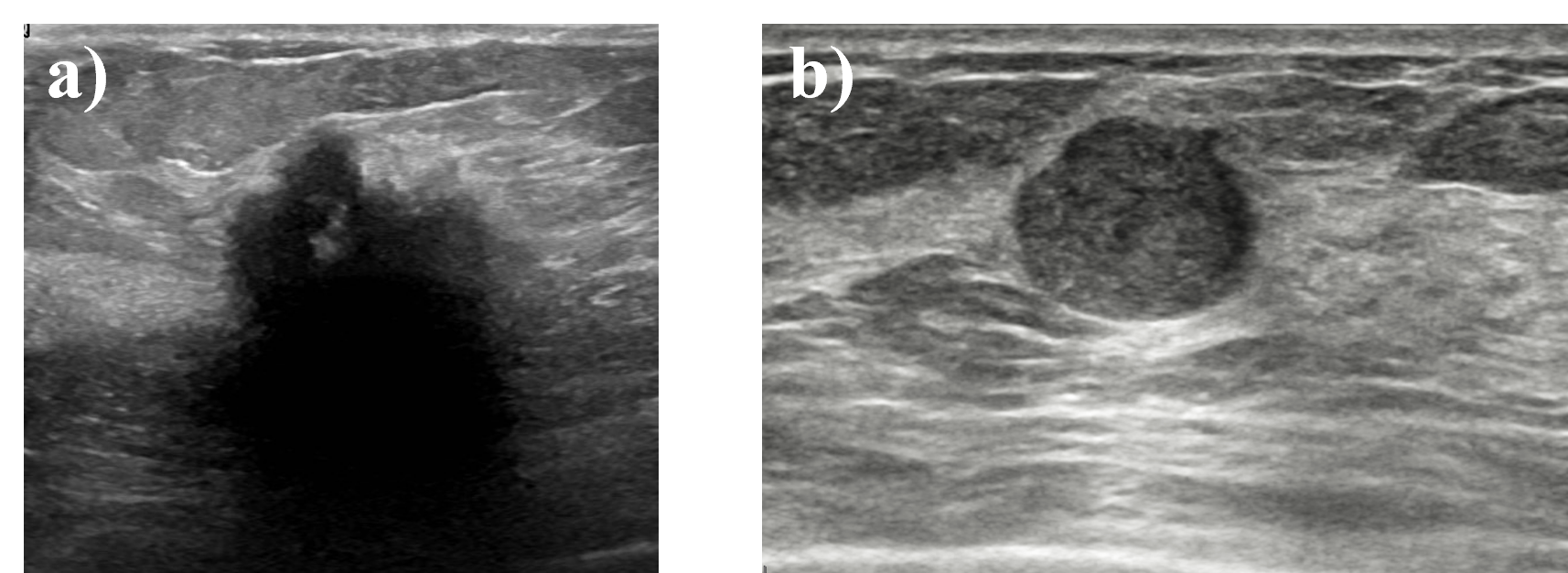}
	\end{center}
	\caption{US images presenting breast masses with a) poorly defined shape due to the shadowing artifact and b) with a 
	well-defined contour.}
	\label{f1}
\end{figure}

\begin{figure*}[t]
	\begin{center}
		\includegraphics[width=0.9\linewidth]{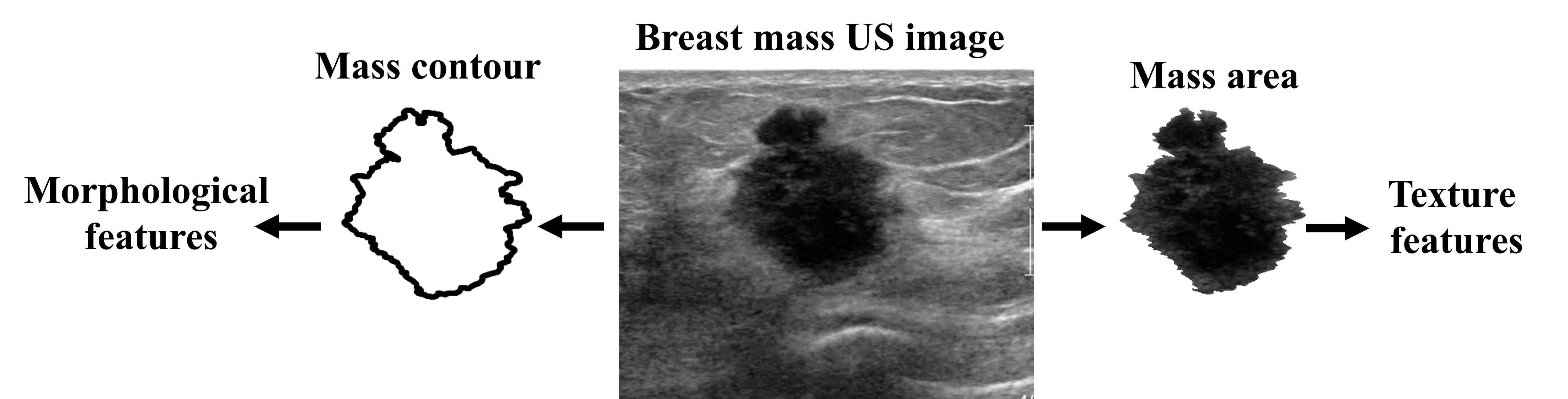}
	\end{center}
	\caption{Scheme presenting the extraction of the morphological and texture features from breast mass US images.}
	\label{f2}
\end{figure*}

\section{Methods}

\subsection{Dataset}

To develop and evaluate the proposed approach, we used breast mass US images collected from three publicly available datasets: BUSI, RODTOOK and UDIAT \cite{al-dhabyani_dataset_2020,rodtook_automatic_2018,yap_automated_2018,yap_breast_2018}. The datasets were processed to remove US images that included scanner annotations within the breast mass regions, which could impact the estimation of the texture features. The dataset after the filtration contained 746 breast mass US images, 302 malignant and 444 benign.  8-fold cross-validation was used to assess the implemented techniques. For each fold, $1/8$ and $7/8$ of the dataset were used for the testing and training, respectively. Additionally, the training set was divided with a 50\%/50\%  split into the development training set and the meta-training set. The development training set was used to train the standard classifiers while the meta-training set was utilized to train the meta-network to perform classifier selection. All sets were balanced to include the same ratio of the US images from each public dataset as well as the same ratio of the malignant and benign breast masses.

\subsection{Standard classifiers}

Following the work of Flores et al., we developed two standard classifiers to differentiate malignant and benign breast masses in US images \cite{flores2015improving}. The first classifier was based on shape parameters. In this case, the following 15 morphological features were determined based on the mass boundary outlines: depth-to-width ratio, mass area, circularity, roundness, normalized residual value, overlap ratio, convexity, orientation, long axis to short axis ratio, elliptic normalized skeleton, elliptic normalized circumference, mean of normalized radial length (NRL), standard deviation of NRL, area ratio and contour roughness \cite{flores2015improving,alvarenga2010assessing,gomez2020assessment}.  The second classifier was trained with the texture features calculated using the gray-level co-occurrence matrix (GLCM) technique. We determined the following GLCM statistics based on the breast mass areas: contrast, correlation, energy, variance, maximum probability and auto-correlation \cite{gomez2012analysis}. Statistics were computed for two quantization levels (4 and 16), four orientations (0$^{\circ}$, 45$^{\circ}$, 90$^{\circ}$ and 135$^{\circ}$) and two distances~(1 and 5 pixels), resulting in 112 texture features. Extraction of the features is illustrated in Fig.~\ref{f2}. Shape features were calculated in Python while the texture features were computed in Matlab (The MathWorks, Inc., USA). 

Logistic regression algorithm with the L1 loss was applied for the binary classification of malignant and benign breast mass. For each cross-validation fold, we used the development dataset to separately train the classifiers based on the shape and texture features. We employed class weights that were inversely proportional to the class frequencies in the training set to address the class imbalance problem. Additionally, based on the development training set the median and interquartile range were determined for each feature and used for feature scaling both in the case of the development training data and the samples from the test set and the meta-learning training set. 

\subsection{Meta-learning}

\begin{figure*}[]
	\begin{center}
		\includegraphics[width=0.7\linewidth]{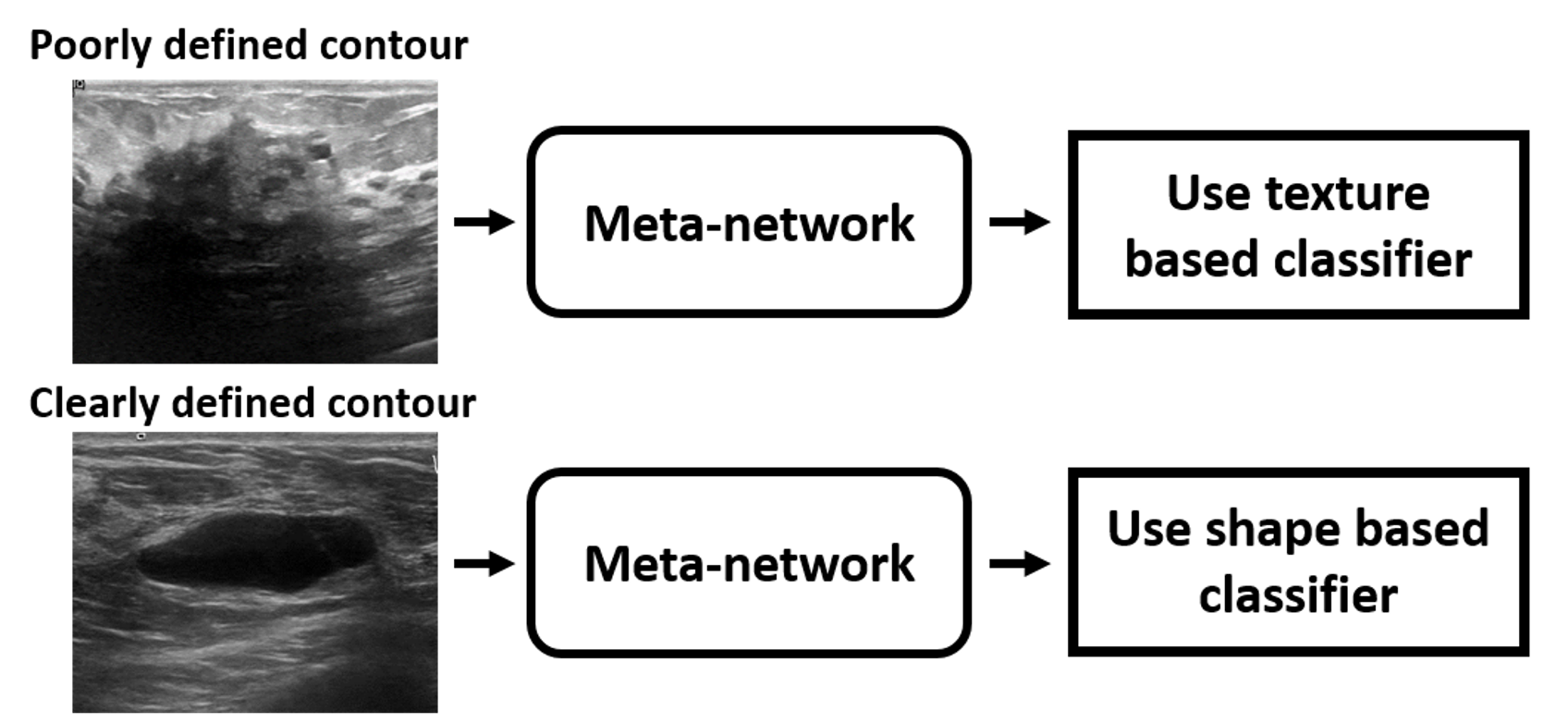}
	\end{center}
	\caption{The meta-network was developed to indicate which standard classifier should be applied to classify the input breast mass US image. We expect that the meta-network will recommend the standard classifier based on input US image apperance.}
	\label{f3}
\end{figure*}

For each cross-validation fold, the classifiers trained on the development training set were evaluated on the meta-learning training set to determine the reference for the training of the meta-network. For each US image from the meta-learning training set we determined the classification error with the following equation: 

\begin{equation}
    e = |p-c|,
\end{equation}

\noindent where $p$ is the probability of mass malignancy outputted by the classifier and $c$ stands for the breast mass class (0 for benign and 1 for malignant). Classification error approaches 0 when the classifier is accurate and 1 otherwise. For each US image from the meta-learning training set we selected the better performing standard method in respect to the classification error. Next, the meta-network was trained to output the better performing standard classifier for each US image, which corresponded to the binary classification setting. Our approach is illustrated in Fig. \ref{f3}, we expect that the meta-network will learn to recommend the more accurate classifier for the particular US image based on particular image characteristics.

Meta-network was trained using the binary cross-entropy loss function.  EfficientNetV2M convolutional neural network pre-trained on the ImageNet dataset served as the backbone for the meta-network \cite{tan2021efficientnetv2}. The original dense layer was replaced by a dense layer equipped with the sigmoid activation function suitable for the binary classification problem. Gray-scale US images were converted to RGB and processed in the same way as the original ImageNet data used for the pre-training. Moreover, US images were cropped using regions of interest provided by the dataset creators and resized to the target image size of 256x256. Fine-tuning with the Adam optimizer and learning rate of 0.0001 was applied to train the meta-network. Calculations were performed in TensorFlow \cite{abadi2016tensorflow}.

\subsection{Evaluation}

For each cross-validation fold, the morphological and texture based classifiers trained on the development training set were evaluated on the test set. Additionally, we used the meta-network trained on the meta-learning training set to recommend the standard classifier for each test US image. To evaluate the implemented methods, we used  the area under the receiver operating characteristic curve (AUC) and accuracy. Additionally, we evaluated the performance of an oracle that always recommend the better performing classifier. This approach corresponds to the best performance achievable with the classifiers based on the morphological and texture features in our meta-learning framework.

\section{Results and Discussion}

\begin{figure}[]
	\begin{center}
		\includegraphics[width=0.7\linewidth]{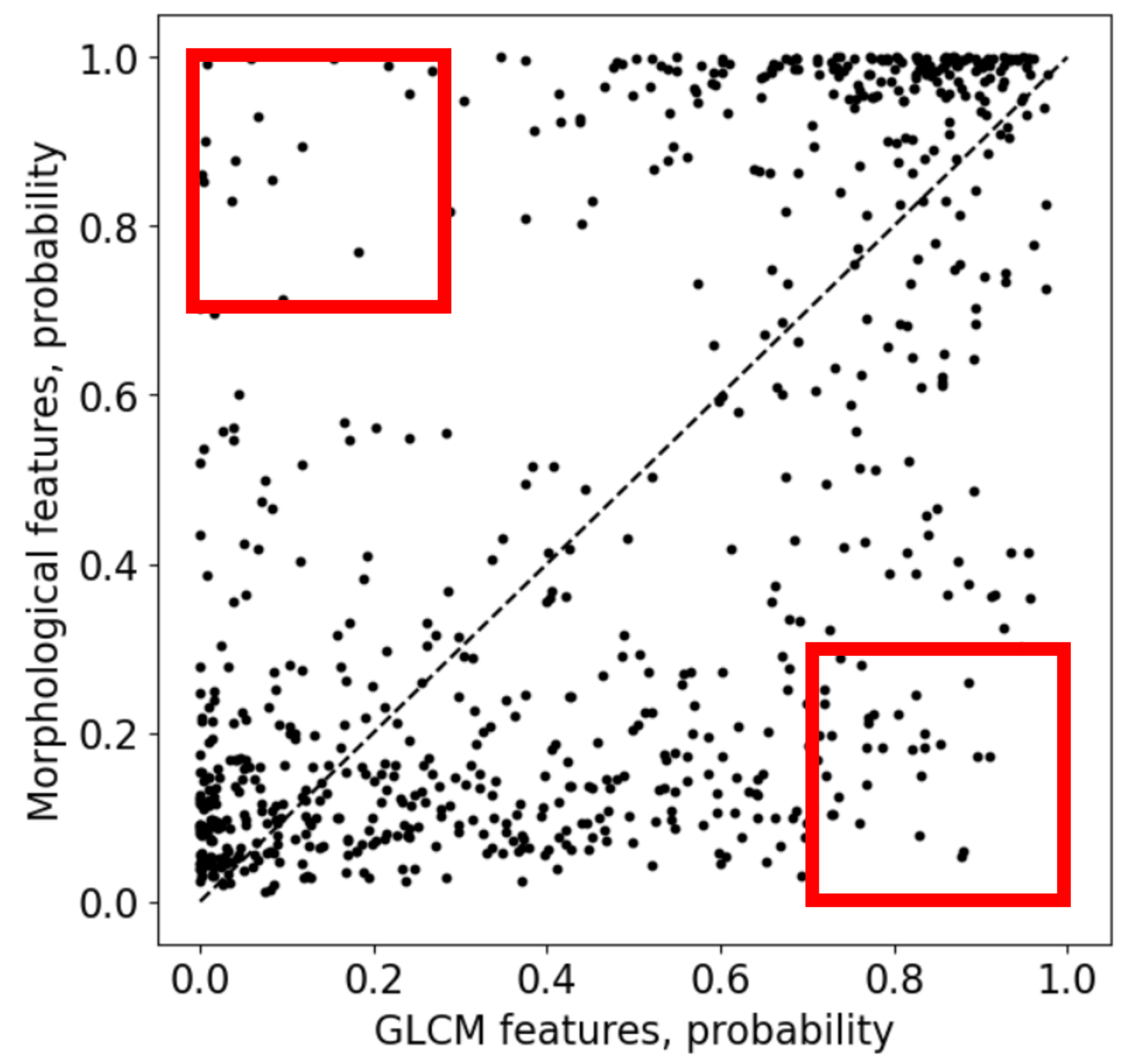}
	\end{center}
	\caption{The agreement between the outputs (probabilities of mass malignancy) of the shape and texture based classifiers determined on the entire data via the 8-fold cross-validation. Red boxes indicate the test cases for which the classifiers disagreed.}
	\label{f4}
\end{figure}

Fig. \ref{f4} presents the relationship between the outputs (probabilities of malignancy) of the morphological and texture based classifiers determined for the entire dataset.  \cite{antropova2017deep}. The correlation coefficient between the outputs was equal to 0.64, which shows that the agreement between the classifiers was fairly strong, but not perfect. In Fig. \ref{f4}, we outlined with the red boxes the cases for which the classifiers strongly disagreed in the assigned probabilities of malignancy. This result suggests that for some breast masses only one of the classifiers should be preferred.

Table 1 summarizes the classification performance obtained for the investigated techniques. The morphological feature based classifier achieved higher AUC value, 0.93, than the texture based method, 0.88. This result confirms the findings of Flores et al. who reported that the morphological features are the better performing features for the  breast mass differentiation \cite{flores2015improving}. The meta-learning based classifier selection resulted in better performance than for the individual standard classifiers. In this case, we achieved AUC and accuracy of 0.95 and 0.91, respectively. Table 1 also presents the performance of an oracle that always recommend the better performing classifier for the test image. In this case, the obtained AUC and accuracy were equal to 0.99 and 0.96, respectively. This result shows that the meta-network did not select the better performing classifier for each US image. Moreover, the accuracy of 0.96 indicates that the perfect classification was not achievable on our dataset with the combination of the shape and texture based classifiers. 

As far as we know, in our work meta-learning was used for the first time to recommend suitable classifiers for the breast mass classification in US. Flores et al. compared the performance of the morphological and texture features for breast mass differentiation \cite{flores2015improving}. However, authors did not investigate the agreement between the classifiers or the reasons for which the classifiers fail to produce accurate predictions. Conclusions of the authors were based solely on the obtained AUC values. In contrary, we presented that the proposed meta-learning technique can be used to successfully select the better performing standard classifier for input US images. Compared to the shape based classifier, with the proposed meta-learning based approach we could increase the AUC value from 0.93 to 0.95.

The proposed approach partially addresses one of the important problems of the deep neural networks, which is the lack of the interpretability. In this study, we did not develop a neural network to perform the diagnosis, but to improve the performance of the standard well-known classification methods. In practice, the deep learning methods may fail on new data due to the data-shifts or adversarial attacks \cite{rabanser2019failing}. However, in the case of our approach, the classification is solely based on the handcrafted features. Even if the meta-network fail to recommend the better performing algorithm for the input US image, still one of the standard classifiers will be used for the analysis, which should grant a certain level of robustness.

In future, we plan to conduct additional experiments to better illustrate the proposed meta-learning based approach. First, we would like to incorporate other classification methods into our framework. For example, it would be interesting to include a classifier based on quantitative US parameters, like the backscatter coefficient \cite{oelze2016review}. Second, we plan to utilize techniques that can be used to generate saliency maps to understand where the convolutional meta-network is focusing in the input US image to output the recommendation \cite{zhou2016learning}.  Third, we would like to improve our framework and take into account the cases that were wrongly assessed by both classifiers. Presence of such cases should be considered when designing the reference for the training of the meta-network.   

\begin{table}[]
\begin{center}

        \caption{Breast mass differentiation performance determined for the standard methods and the approaches utilizing the meta-learning based classifier selection. }
        \label{t1}

\scalebox{0.75}{
\begin{tabular}{|c|c|c|}
\hline
    
    Method & AUC & Accuracy  \\
                  \hline \hline
                  
    GLCM features & 0.88 & 0.81  \\  \hline     
    
    Morphological features & 0.93 & 0.88  \\  \hline     

    Meta-learning  & 0.95 & 0.91  \\  \hline     
    
    Oracle & 0.99 & 0.96  \\  \hline       
                  
\end{tabular}} 
\end{center}
\end{table}

\section{Conclusion}

In this preliminary study, we investigated the usefulness of the meta-learning techniques in the context of  breast mass classification in ultrasound. We developed a meta-network that could recommend the most suitable standard classifier for particular breast mass US image. Obtained results demonstrated that meta-learning can be used to improve the performance of the standard classifiers based on handcrafted features. 

\section*{Conflicts of interest}

The authors do not have any conflicts of interest to disclosure. 

\section*{Acknowledgement}

This work was supported by the National Centre for Research and Development of Poland (grant number INFOSTRATEG-I/0042/2021). 

\bibliographystyle{IEEEtran}
\bibliography{IEEEabrv,references}

\begin{thebibliography}{10}
\providecommand{\url}[1]{#1}
\csname url@samestyle\endcsname
\providecommand{\newblock}{\relax}
\providecommand{\bibinfo}[2]{#2}
\providecommand{\BIBentrySTDinterwordspacing}{\spaceskip=0pt\relax}
\providecommand{\BIBentryALTinterwordstretchfactor}{4}
\providecommand{\BIBentryALTinterwordspacing}{\spaceskip=\fontdimen2\font plus
\BIBentryALTinterwordstretchfactor\fontdimen3\font minus
  \fontdimen4\font\relax}
\providecommand{\BIBforeignlanguage}[2]{{%
\expandafter\ifx\csname l@#1\endcsname\relax
\typeout{** WARNING: IEEEtran.bst: No hyphenation pattern has been}%
\typeout{** loaded for the language `#1'. Using the pattern for}%
\typeout{** the default language instead.}%
\else
\language=\csname l@#1\endcsname
\fi
#2}}
\providecommand{\BIBdecl}{\relax}
\BIBdecl

\bibitem{houssein2021deep}
E.~H. Houssein, M.~M. Emam, A.~A. Ali, and P.~N. Suganthan, ``Deep and machine
  learning techniques for medical imaging-based breast cancer: A comprehensive
  review,'' \emph{Expert Systems with Applications}, vol. 167, p. 114161, 2021.

\bibitem{abdullah2021review}
T.~A. Abdullah, M.~S.~M. Zahid, and W.~Ali, ``A review of interpretable ml in
  healthcare: Taxonomy, applications, challenges, and future directions,''
  \emph{Symmetry}, vol.~13, no.~12, p. 2439, 2021.

\bibitem{flores2015improving}
W.~G. Flores, W.~C. de~Albuquerque~Pereira, and A.~F.~C. Infantosi, ``Improving
  classification performance of breast lesions on ultrasonography,''
  \emph{Pattern Recognition}, vol.~48, no.~4, pp. 1125--1136, 2015.

\bibitem{byra2019quantitative}
M.~Byra, L.~Wan, J.~H. Wong, J.~Du, S.~B. Shah, M.~P. Andre, and E.~Y. Chang,
  ``Quantitative ultrasound and b-mode image texture features correlate with
  collagen and myelin content in human ulnar nerve fascicles,''
  \emph{Ultrasound in medicine \& biology}, vol.~45, no.~7, pp. 1830--1840,
  2019.

\bibitem{khan2020literature}
I.~Khan, X.~Zhang, M.~Rehman, and R.~Ali, ``A literature survey and empirical
  study of meta-learning for classifier selection,'' \emph{IEEE Access},
  vol.~8, pp. 10\,262--10\,281, 2020.

\bibitem{al-dhabyani_dataset_2020}
W.~Al-Dhabyani, M.~Gomaa, H.~Khaled, and A.~Fahmy,
  ``\BIBforeignlanguage{en}{Dataset of breast ultrasound images},''
  \emph{\BIBforeignlanguage{en}{Data in Brief}}, vol.~28, p. 104863, Feb. 2020.

\bibitem{rodtook_automatic_2018}
A.~Rodtook, K.~Kirimasthong, W.~Lohitvisate, and S.~S. Makhanov,
  ``\BIBforeignlanguage{en}{Automatic initialization of active contours and
  level set method in ultrasound images of breast abnormalities},''
  \emph{\BIBforeignlanguage{en}{Pattern Recognition}}, vol.~79, pp. 172--182,
  Jul. 2018.

\bibitem{yap_automated_2018}
M.~H. Yap, G.~Pons, J.~Marti, S.~Ganau, M.~Sentis, R.~Zwiggelaar, A.~K.
  Davison, R.~Marti, n.~Moi Hoon~Yap, G.~Pons, J.~Marti, S.~Ganau, M.~Sentis,
  R.~Zwiggelaar, A.~K. Davison, and R.~Marti,
  ``\BIBforeignlanguage{eng}{Automated {Breast} {Ultrasound} {Lesions}
  {Detection} {Using} {Convolutional} {Neural} {Networks}},''
  \emph{\BIBforeignlanguage{eng}{IEEE Journal of Biomedical and Health
  Informatics}}, vol.~22, no.~4, pp. 1218--1226, 2018.

\bibitem{yap_breast_2018}
M.~H. Yap, M.~Goyal, F.~M. Osman, R.~Martí, E.~Denton, A.~Juette, and
  R.~Zwiggelaar, ``Breast ultrasound lesions recognition: End-to-end deep
  learning approaches,'' \emph{Journal of Medical Imaging}, vol.~6, no.~1, p.
  011007, Oct. 2018, publisher: International Society for Optics and Photonics.

\bibitem{alvarenga2010assessing}
A.~V. Alvarenga, A.~F.~C. Infantosi, W.~C.~A. Pereira, and C.~M. Azevedo,
  ``Assessing the performance of morphological parameters in distinguishing
  breast tumors on ultrasound images,'' \emph{Medical engineering \& physics},
  vol.~32, no.~1, pp. 49--56, 2010.

\bibitem{gomez2020assessment}
W.~G{\'o}mez-Flores and J.~Hern{\'a}ndez-L{\'o}pez, ``Assessment of the
  invariance and discriminant power of morphological features under geometric
  transformations for breast tumor classification,'' \emph{Computer methods and
  programs in biomedicine}, vol. 185, p. 105173, 2020.

\bibitem{gomez2012analysis}
W.~G{\'o}mez, W.~C.~A. Pereira, and A.~F.~C. Infantosi, ``Analysis of
  co-occurrence texture statistics as a function of gray-level quantization for
  classifying breast ultrasound,'' \emph{IEEE transactions on medical imaging},
  vol.~31, no.~10, pp. 1889--1899, 2012.

\bibitem{tan2021efficientnetv2}
M.~Tan and Q.~Le, ``Efficientnetv2: Smaller models and faster training,'' in
  \emph{International Conference on Machine Learning}.\hskip 1em plus 0.5em
  minus 0.4em\relax PMLR, 2021, pp. 10\,096--10\,106.

\bibitem{abadi2016tensorflow}
M.~Abadi, P.~Barham, J.~Chen, Z.~Chen, A.~Davis, J.~Dean, M.~Devin,
  S.~Ghemawat, G.~Irving, M.~Isard \emph{et~al.}, ``$\{$TensorFlow$\}$: a
  system for $\{$Large-Scale$\}$ machine learning,'' in \emph{12th USENIX
  symposium on operating systems design and implementation (OSDI 16)}, 2016,
  pp. 265--283.

\bibitem{antropova2017deep}
N.~Antropova, B.~Q. Huynh, and M.~L. Giger, ``A deep feature fusion methodology
  for breast cancer diagnosis demonstrated on three imaging modality
  datasets,'' \emph{Medical physics}, vol.~44, no.~10, pp. 5162--5171, 2017.

\bibitem{rabanser2019failing}
S.~Rabanser, S.~G{\"u}nnemann, and Z.~Lipton, ``Failing loudly: An empirical
  study of methods for detecting dataset shift,'' \emph{Advances in Neural
  Information Processing Systems}, vol.~32, 2019.

\bibitem{oelze2016review}
M.~L. Oelze and J.~Mamou, ``Review of quantitative ultrasound: Envelope
  statistics and backscatter coefficient imaging and contributions to
  diagnostic ultrasound,'' \emph{IEEE transactions on ultrasonics,
  ferroelectrics, and frequency control}, vol.~63, no.~2, pp. 336--351, 2016.

\bibitem{zhou2016learning}
B.~Zhou, A.~Khosla, A.~Lapedriza, A.~Oliva, and A.~Torralba, ``Learning deep
  features for discriminative localization,'' in \emph{Proceedings of the IEEE
  conference on computer vision and pattern recognition}, 2016, pp. 2921--2929.

\end{thebibliography}

\end{document}